\documentstyle[11pt,moriond,epsfig]{article}

\def\Journal#1#2#3#4{{#1} {\bf #2}, #3 (#4)}

\def\PLB{{\em Phys. Lett.} B}
\def\PRL{\em Phys. Rev. Lett.}
\def\PRD{{\em Phys. Rev.} D}

\newcommand{\nullchar}{\hspace*{0ex}}
\newcommand{\gtap}{{\raise.3ex\hbox{$>$\kern-.75em\lower1ex\hbox{$\sim$}}}}
\newcommand{\ltap}{{\raise.3ex\hbox{$<$\kern-.75em\lower1ex\hbox{$\sim$}}}}
\def\alt{\ltap}
\def\agt{\gtap}

\def\cb{\bar{\chi}}
\def\cbe{\bar{\chi}_{\rm eff}}

\begin{document}
\vspace*{4cm}
\title{A SUPERSYMMETRIC SOLUTION \\TO THE BOTTOM-QUARK CROSS SECTION ANOMALY}

\author{\bf ZACK SULLIVAN}

\address{High Energy Physics Division, Argonne National Laboratory, Argonne,
Illinois 60439}

\maketitle\abstracts{In this talk, I describe a supersymmetric solution to
the long-standing discrepancy between the bottom-quark production cross
section and predictions of perturbative quantum chromodynamics.  Pair
production of light gluinos, of mass 12--16 GeV, with two-body decays into
bottom quarks and bottom squarks, of mass 2--5.5 GeV, yields the correct
normalizations and shapes of the measured bottom-quark distributions.  One
prediction of this scenario is that like-sign $B$ mesons, $B^+B^+$ and
$B^-B^-$, should be produced with a measurable rate at the next run of the
Fermilab Tevatron Collider.}

\section{Introduction}
\vspace*{-1.5ex}
\indent\indent The cross section for bottom-quark production is measured at
hadron and photon colliders to be about a factor of 2 above the
expectations of next-to-leading order (NLO) calculations in perturbative
quantum chromodynamics (QCD).\cite{expxsec} Despite more than ten years of
effort, this discrepancy has resisted satisfactory resolution within the
standard model (SM) of particle physics.\cite{qcdrev} This is surprising
because the mass of the bottom quark sets a scale at which other
perturbative QCD calculations are reliable.  While additional higher-order
QCD effects in production or fragmentation may solve part or all of the
puzzle, a reasonable question to ask is whether this anomaly is a hint of
``new physics''.

In a recent Letter,\cite{zspaper} we explore an explanation of the
discrepancy within the context of the minimal supersymmetric standard model
(MSSM). We postulate the existence of a light bottom squark
$\tilde b$ (mass $\simeq 2$--5.5 GeV) and a relatively light gluino $\tilde
g$ (mass $\simeq 12$--16 GeV) that decays with $100\%$ branching fraction
to $b$ and $\tilde b$.  The masses of these particles are constrained by
fits to several different experiments as described below.  The $\tilde b$
may either be long-lived or it may decay via $R$-parity-violating
interactions into a pair of hadronic jets.  We obtain good agreement with
the magnitude and shape of the measured distributions of bottom-quark
production at UA1 and the Fermilab Tevatron.  We also make several
predictions, and point out a ``golden channel'' of like-sign $B$ mesons,
$B^+B^+$ or $B^-B^-$, that may either be observed, or whose absence will
rule out this scenario, at run II of the Tevatron.

Our assumptions are consistent with all experimental constraints on the
masses and couplings of supersymmetric particles.$^{4-7}$ The tree-level
coupling of the light $\tilde{b}_1$ to the $Z$ boson $g_{Z\tilde b_1\tilde
b_1}\sim (T_3\sin^2\theta_{\tilde{b}} - Q_{\tilde b} \sin^2\theta_W)$.
Hence, if $\sin\theta_{\tilde{b}} \simeq 0.38$, $\tilde{b}_1$ approximately
decouples from the $Z$, which leads to good agreement with the $Z$-peak
observables.\cite{CHWW} The couplings $g_{Z\tilde b_1\tilde b_2}$ and
$g_{Z\tilde b_2\tilde b_2}$ survive, but are irrelevant as long as
$m_{\tilde b_2}\,\agt\, 200$ GeV.  Production of $\tilde{b}_1$ pairs via
virtual photons is a factor of 2--4 smaller than the best bound from
LEP.\cite{delphilim} Bottom squarks make a tiny ($\sim2\%$) contribution to
$e^+e^-\to$ hadrons.  Thus, despite the improved 6--10\% measurement of $R$
by the BES Collaboration~\cite{besii} presented at this meeting, there is
no sensitivity to this resonance.  Spin-1/2 quarks are produced in $e^+e^-$
annihilations with an angular distribution of $(1+\alpha\cos^2\theta)$ and
$\alpha=1$.  The bottom squark appears as an effective $\alpha\simeq 0.92$.
We refit the angular distribution measured by the CELLO
Collaboration,\cite{CELLO} and find it is consistent with the production of
a single pair of charge-1/3 squarks along with five flavors of
quark-antiquark pairs.

\vspace*{-1.5ex}
\section{Comparison with Data}
\vspace*{-1.5ex}
\indent\indent Because the excess production rate is observed in all
bottom-quark decay channels and distributions, any solution will
necessarily involve additional production of bottom quarks.  In our
scenario, light gluinos are dominantly produced by gluon fusion ($g g \to
\tilde g \tilde g$) at Tevatron energies.  As long as $m_{\tilde g} > m_b +
m_{\tilde b}$, the $\tilde g$ decays promptly to $b + \tilde b$.
\begin{figure}[tb]
\begin{center}\vspace*{-3ex}
\psfig{figure=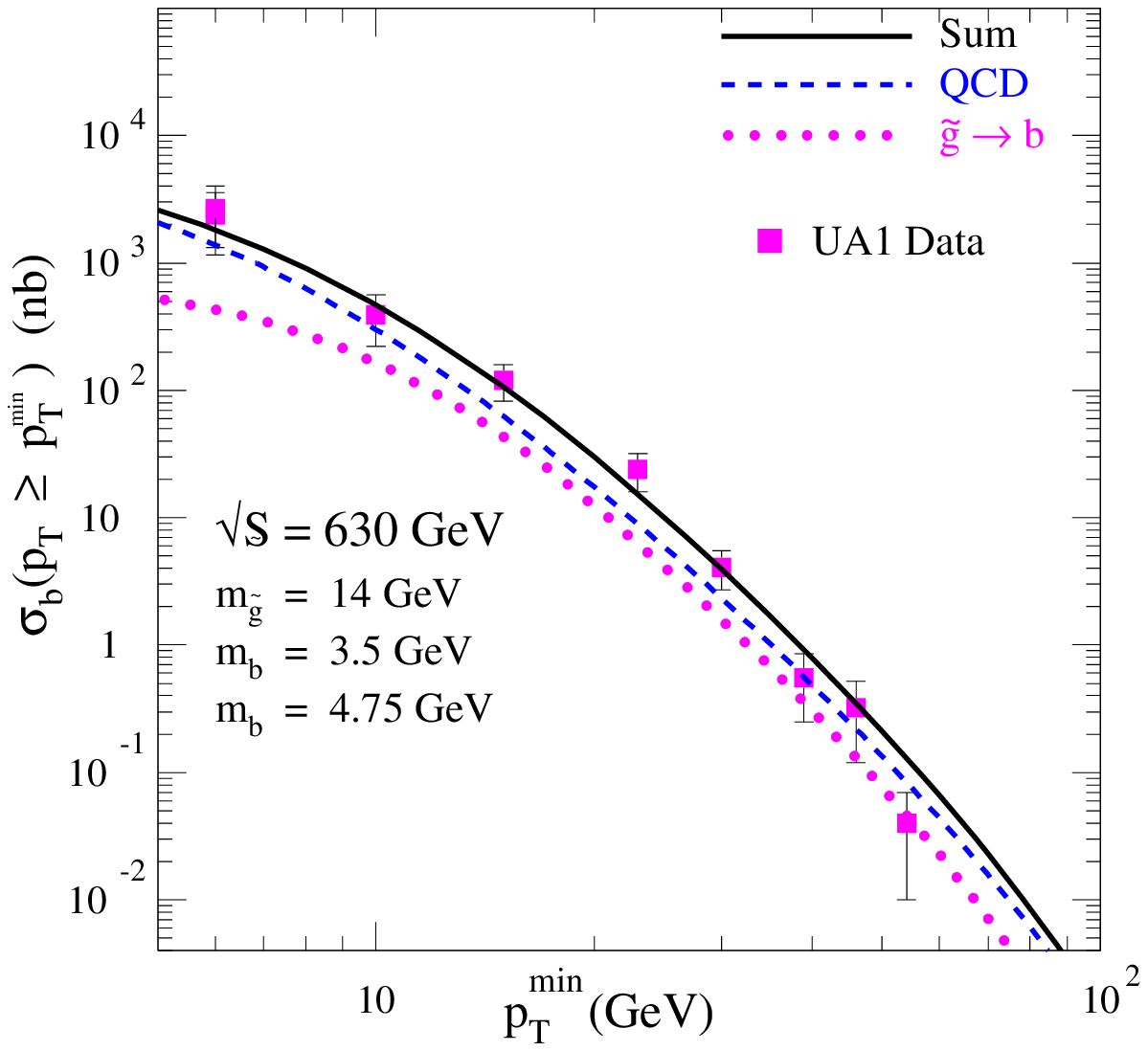,width=2.8in}\hspace*{2em}\psfig{figure=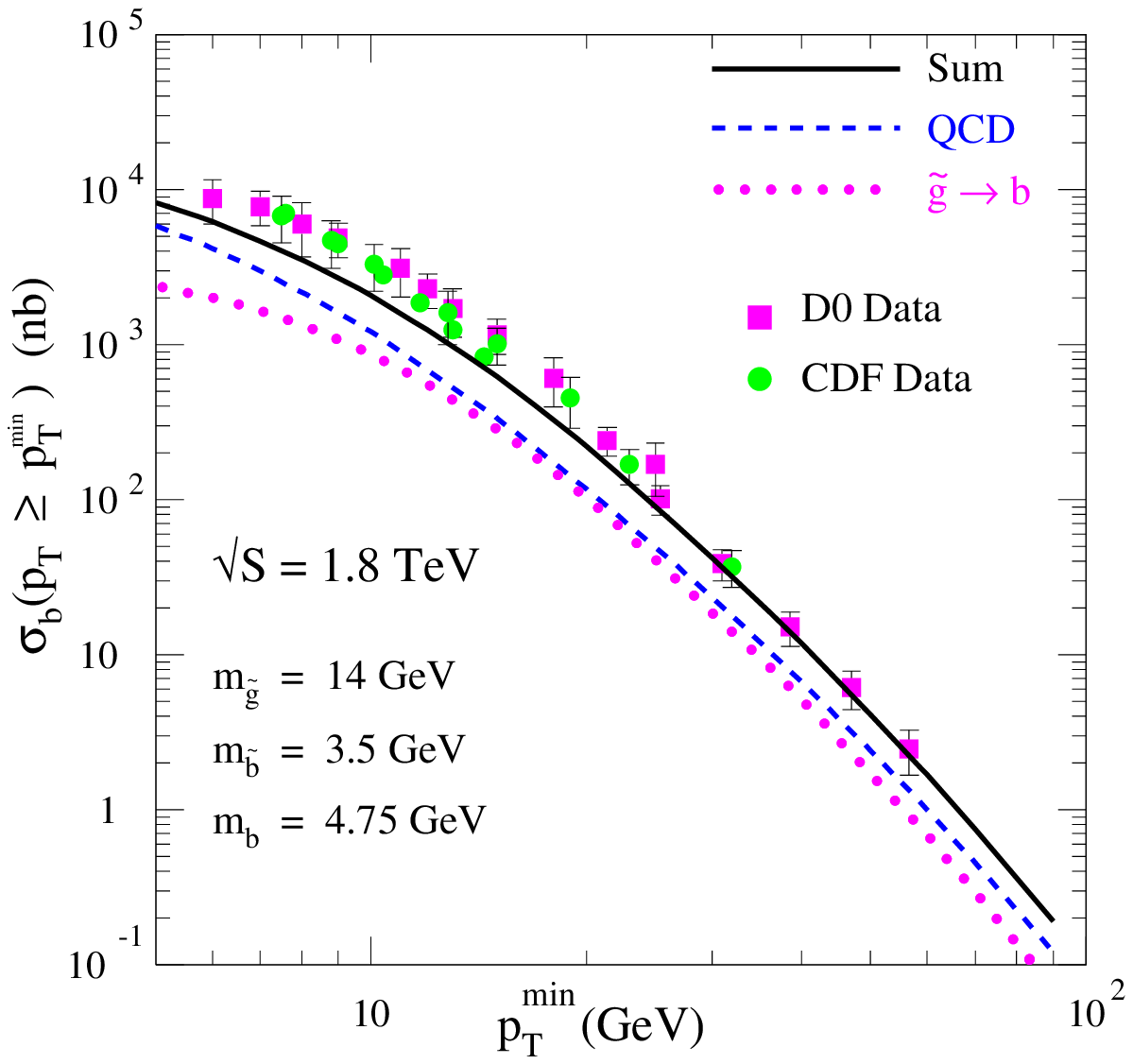,width=2.8in}
\end{center}
\caption{Bottom-quark cross section at UA1 ($\sqrt{S} =630$ GeV) and the
Tevatron ($\sqrt{S}=1.8$ TeV) for $p_{Tb}>p_{Tb}^{\rm min}$ with
$m_{\tilde{g}} = 14$ GeV and $m_{\tilde{b}} = 3.5$ GeV (solid); NLO QCD
prediction (dashed); SUSY contribution (dotted).\hfill\nullchar}
\label{fig:fig1}
\end{figure}
In Fig.~1 we show the integrated $p_{Tb}$ distribution of the $b$ quarks as
measured at UA1~\cite{uaone} and the Tevatron.\cite{expxsec} For comparison
we plot the NLO cross section with CTEQ4M PDF's, $m_b =$ 4.75 GeV, scale
$\mu=\sqrt{m_b^2 + p_{Tb}^2}$.\cite{calcs} We show separately the effect of
$\tilde g$ production, followed by $\tilde g \rightarrow b + \tilde b$, for
$m_{\tilde g} = $14 GeV and $m_{\tilde b} =$ 3.5 GeV.  We compute the
$\tilde g$-pair cross section from the leading order (LO) matrix element
with NLO PDF's, $\mu=\sqrt{m^2_{\tilde g}+p^2_{T\tilde g}}$, a two-loop
$\alpha_s$, and use a $K$-factor of 1.9.\cite{calcs} The SUSY-QCD
corrections to $b \bar{b}$ production are not yet available.\cite{bbsusy}

A gluino of mass $m_{\tilde g} \simeq$ 12--16 GeV is necessary give the
correct magnitude of the cross section.  The $\tilde g$ decays produce
$p_{Tb}$ spectra that are enhanced primarily in the neighborhood of
$p_{Tb}^{\rm min} \simeq m_{\tilde g}$, exactly where the data show the
most prominent enhancement above the QCD expectation.  Larger values of
$m_{\tilde g}$ yield too little cross section to be of interest, but are
not ruled out.  The interesting values of $m_{\tilde b}$ and $m_{\tilde g}$
are correlated; after selections on $p_{Tb}^{\rm min}$, large values of
$m_{\tilde b}$ reduce the cross section and lead to shapes of the $p_{Tb}$
distribution that agree less well with the data.

After the contributions of the NLO QCD and SUSY components are added (solid
curve in Fig.~1), the magnitude of the bottom-quark cross section and the
shape of the integrated $p^{\rm min}_{Tb}$ distribution are described well.
A theoretical uncertainty of roughly $\pm 30\%$ may be assigned to the
final solid curve, associated with variation of the $b$ mass, the scale,
and the parton distributions.  The SUSY process produces bottom quarks in a
four-body final state and thus their momentum correlations are different
from those of QCD.  Angular correlations between muons that arise from
decays of $b$'s have been measured.\cite{cdfmix,muonexp} Examining the
angular correlations between $b$'s in the SUSY case we find they are nearly
indistinguishable from those of QCD once experimental cuts are applied.

\vspace*{-1.5ex}
\section{Effects on $B^0$--$\bar B^0$ Mixing}
\vspace*{-1.5ex} \indent\indent Since the $\tilde g$ is a Majorana
particle, its decay can yield either quarks or antiquarks.  Given the
kinematic cuts applied at hadron colliders, gluino pair production and
subsequent decay to $b$'s will generate a number of $b b$ and $\bar b \bar
b$ pairs equal to the number of $b \bar b$ final states.\cite{zspaper} This
leads to the ``golden signature'' of like-sign $B$ mesons, $B^+B^+$ and
$B^-B^-$.  If these do not appear in the run II data, then this scenario
may be ruled out.

We predict there will be an increase of like-sign leptons in the final
state after semi-leptonic decays of the $b$ and $\bar b$ quarks.  This
increase could be confused with an enhanced rate of $B^0$--$\bar B^0$ mixing.
Time-integrated mixing analyses of lepton pairs determine the quantity $\cb
= f_d \chi_d + f_s \chi_s$, where $f_d$ and $f_s$ are the fractions of
$B^0_d$ and $B^0_s$ hadrons, respectively, in the sample of semi-leptonic
$B$ decays, and $\chi_f$ is the time-integrated mixing probability for
$B^0_f$.  The quantity $2\cb (1-\cb)$ is the fraction of $b\bar b$ pairs
that decay as like-sign $b$'s.  Our SUSY mechanism can be incorporated by
introducing $\cbe$ such that $2\cbe (1-\cbe)=[2\cb (1-\cb) + G/2]/(1+G)$,
where $G$ is the ratio of SUSY and QCD bottom-quark cross sections after
cuts.  Hadron colliders measure
\begin{equation}
\cbe=\frac{\cb}{\sqrt{1+G}} +{1\over 2}\left[1-
	\frac{1}{\sqrt{1+G}}\right] \;.
\end{equation}

To estimate $\cbe$, we assume that the world average value $\cb = 0.118 \pm
0.005$ \cite{pdg} represents the contribution from only the pure QCD
component.  We determine $\cbe$ in the region of phase space where the
measurement is made,\cite{cdfmix} with both final $b$'s having $p_T$ of at
least 6.5 GeV and rapidity $| y_b | \leq 1$.  For gluino masses of
$m_{\tilde g} =$ 14 and 16 GeV, we obtain $\cbe = 0.17\pm 0.02$ and
$0.16\pm 0.02$, respectively.  There is an additional uncertainty of $\pm
0.02$ from the lack of a NLO calculation of $\tilde g\to b\tilde b$
distributions.  Our expectations may be compared with the CDF
Collaboration's published value $\cbe = 0.131 \pm 0.02 \pm
0.016$.\cite{cdfmix} Values of $m_{\tilde g} > 12$ GeV lead to a calculated
$\cbe$ that is consistent with the measured value within experimental and
theoretical \nolinebreak uncertainties.

\vspace*{-1.5ex}
\section{Additional Implications}
\vspace*{-1.5ex}
\indent\indent In the standard model, a global fit to all observables
provides an indirect measurement of $\alpha_s(M_Z) \simeq 0.119 \pm
0.006$.\cite{pdg} A light $\tilde g$ with mass about 15 GeV and a light
$\tilde b$ modify the QCD $\beta$ function.  Thus, experiments performed
below $m_{\tilde g}$ would predict $\alpha_s(M_Z)=0.125$.  This is within
the range of experimental uncertainty, and in better agreement with some
measurements at LEP.  Light gluinos are also helpful in improving gauge
coupling unification by providing a light threshold in the evolution of
$\alpha_s$.  For gluinos of mass several hundred GeV, the strong coupling
at $M_Z$ predicted from unification is somewhat above 0.13.  However, for
gluinos of mass 15 GeV, this prediction becomes $\alpha_s (M_Z) \alt
0.127$, in agreement with the measured value.

If the $\widetilde{b}$ is relatively stable, the $\widetilde{b}$ could pick
up a light $\bar{u}$ or $\bar{d}$ and become a $\widetilde{B}^-$ or
$\widetilde{B}^0$ ``mesino'' with $J = 1/2$.  The mass of the mesino would
be roughly $3$--$7$ GeV for the interval of $\widetilde{b}$ masses we
consider.  The charged mesino could fake a heavy muon if it punches through
the hadron calorimeter, or perhaps act like a heavy $\bar{p}$ --- possibly
detectable with a time-of-flight apparatus.  A long-lived $\widetilde b$ is
not excluded by conventional searches at hadron and lepton colliders.

If $R$ parity is violated, the bottom squark can decay either promptly, or
somewhere outside of the detector.  The CLEO Collaboration~\cite{CLEO} has
constrained a promptly decaying $\tilde b$ with mass 3.5--4.5 GeV with the
decay chain $\tilde b \rightarrow c \em{l} \tilde \nu$ or $\tilde b
\rightarrow c {\em l}$.  Baryon-number-violating decays, however, are
nearly unconstrained.\cite{bhs} In this case, the bottom squark decays to 2
jets with a width~\cite{bhs}
\begin{equation}
\Gamma(\tilde b \to jj)=
\frac{m_{\tilde b}}{2\pi} \sin^2\theta_{\tilde{b}} 
\sum_{j<k} |\lambda^{\prime\prime}_{ij3}|^2 \;.\vspace*{-1.5ex}
\end{equation}
If $m_{\tilde b} = 3.5$ GeV, $\Gamma(\widetilde b \rightarrow i j) = 0.08
|\lambda^{\prime\prime}_{ij3}|^2$ GeV.  Unless all
$\lambda^{\prime\prime}_{ij3}$ are extremely small, the $\widetilde {b}$
will decay quickly and leave soft hadrons in the cone around the $b$-jet.

There appears to be roughly a factor of 2 difference between the QCD
prediction and the $b$ production rate measured at the $ep$ collider HERA
and in photon-photon collisions at LEP.\cite{heralepb} Whether the
existence of light bottom squarks and gluinos in the mass ranges we
consider will produce enough of an excess to explain these experiments is
unknown.  A full NLO study is underway to determine the
effect.\cite{bbsusy}

\vspace*{-1.5ex}
\section*{Acknowledgments}
\vspace*{-1.5ex}
I am indebted to Ed Berger, Brian Harris, David E. Kaplan, Tim Tait, and
Carlos Wagner for their collaboration and insight.  This work is supported
by the U.S. Department of Energy under Contract W-31-109-ENG-38.

\end{document}